\documentclass[12pt]{article}  
\newcommand{\comma}{\, ,}
\newcommand{\period}{\;\; .}

\newcommand{\eq}{\; = \;}
\newcommand{\sep}{\, , \;\;}
\newcommand{\be}{\begin{equation}}
\newcommand{\bd}{\begin{displaymath}}
\newcommand{\ee}{\end{equation}}
\newcommand{\ed}{\end{displaymath}}
\newcommand{\ba}{\begin{eqnarray}}
\newcommand{\ea}{\end{eqnarray}}

\newcommand{\minus}{\! - \!}

\newcommand{\plus}{\! + \!}

\renewcommand{\i}{{\rm i}}
\newcommand{\e}{{\rm e}}
\newcommand{\x}{{x}}
\newcommand{\y}{{y}}
\newcommand{\I}{{\rm I}}
\newcommand{\brk}{ \! \! \! &  \! \! \! &  \! \! \! \!}
\newcommand{\close}{ \! \! \! \! \! \! \! \! \! \!}

\title{Transfer matrix functional relations for the generalized 
$\tau_2(t_q)$ model}

\author{ R.J. Baxter\\
{\protect \small  Mathematical
Sciences Institute}\\
{\protect \small  The Australian National University,
 Canberra, A.C.T. 0200, Australia  }
}

\date{Thursday 4 March 2004}

\begin{document}


\maketitle

\abstract{The $N$-state chiral Potts model in lattice statistical 
mechanics can be obtained as a ``descendant'' of  the six-vertex 
model, via an intermediate ``$Q$'' or ``$\tau_2 (t_q)$'' model.
Here we generalize this to obtain a column-inhomogeneous 
$\tau_2 (t_q)$ model, and derive the functional relations satisfied
by its row-to-row transfer matrix. We do {\em not} need the usual 
chiral Potts
relations between the $N$th powers of the rapidity parameters
$a_p, b_p, c_p, d_p$ of each column. This enables us to readily
consider the case of fixed-spin boundary conditions on the left 
and right-most columns. We thereby re-derive the simple direct 
product structure of the  transfer matrix eigenvalues of this
model, which is closely related to the superintegrable
chiral Potts model with fixed-spin boundary conditions.}




\section*{Introduction}
In a remarkable paper,\cite{BazStrog90}  Bazhanov and Stroganov showed 
in  1990 how the  recently-discovered solvable chiral Potts model 
could be obtained from the six-vertex model by a two-stage process. 
First one looked for a ``$Q$'' or $\tau_2 (t_q)$ 
model whose column-to-column transfer matrix commuted with that of the 
six-vertex model. This turned out to be a spin model on the square 
lattice, each spin taking a given number $N$ of values. Then one 
looked 
for a third model whose row-to-row transfer matrix commuted with that 
of the $\tau_2 (t_q)$ model. This was the $N$-state chiral Potts 
model. Some of this working  was re-presented and extended by Baxter, 
Bazhanov and  Perk.\cite{BBP90,RJB2004}

Here we focus attention on the square lattice $\cal L$ of $L$ columns. 
With row  $i$ we associate a horizontal ``rapidity'' $q_i$. With 
column $j$ we  associate two successive vertical rapidities 
$p_{2j-1}, p_{2j}$, as in Figure \ref{sqlattice} (except for the 
initial six-vertex model, which has only one rapidity line per column,
as we mention below).

\setlength{\unitlength}{1pt}
\begin{figure}[hbt]
\begin{picture}(420,160) (0,0)
\put (60,0) {\line(1,0) {260}}
\put (60,42) {\line(1,0) {260}}
\put (60,84) {\line(1,0) {260}}
\put (60,126) {\line(1,0) {260}}
\multiput(60,0)(52,0){6}{\line(0,1) {126}}
\multiput(60,0)(52,0){6}{\circle*{7}}
\multiput(60,42)(52,0){6}{\circle*{7}}
\multiput(60,84)(52,0){6}{\circle*{7}}
\multiput(60,126)(52,0){6}{\circle*{7}}
\multiput(45,21)(5,0){59}{.}
\multiput(45,63)(5,0){59}{.}
\multiput(45,105)(5,0){59}{.}
\multiput(73,-20)(0,5){32}{.}
\multiput(99,-20)(0,5){32}{.}
\multiput(125,-20)(0,5){32}{.}
\multiput(151,-20)(0,5){32}{.}
\multiput(177,-20)(0,5){32}{.}
\multiput(203,-20)(0,5){32}{.}
\multiput(229,-20)(0,5){32}{.}
\multiput(255,-20)(0,5){32}{.}
\multiput(281,-20)(0,5){32}{.}
\multiput(307,-20)(0,5){32}{.}
\put (345,18) {$q_1$}
\put (345,60) {$q_2$}
\put (345,102) {$q_3$}
\put (71,-31) {$p_1$}
\put (97,-31) {$p_2$}
\put (123,-31) {$p_3$}
\put (149,-31) {$p_4$}
\put (302,-31) {$p_{2L}$}
\put (58,-17) {$1$}
\put (110,-17) {$2$}
\put (162,-17) {$3$}
\put (266,-17) {$L$}
\put (318,-17) {$1$}
\put (155,32) {$i$}
\put (155,74) {$l$}
\put (207,32) {$j$}
\put (207,74) {$k$}
 \end{picture}
\vspace{1.5cm}
 \caption{ The square lattice $\cal L$ of $L$ columns with  cyclic 
boundary   conditions, showing the horizontal  rapidity lines 
$q_1, q_2, \ldots $  and  the vertical rapidity  lines 
$p_1, p_2, \ldots , p_{2L}$~.}
 \label{sqlattice}
\end{figure}
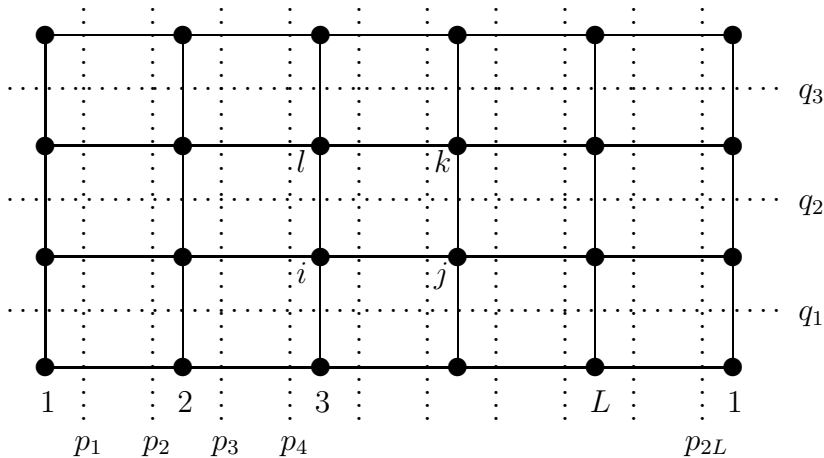

The $p$ and $q$ rapidities are of different types: $q$ is a six-vertex  
model rapidity, specified by  a single complex variable $t_q$, while 
$p$ is a chiral Potts  model 
rapidity, specified by four homogeneous variables $a_p, b_p, c_p, d_p$. 
Because the aim of the earlier papers
was to extablish a link between the six-vertex and homogeneous chiral
Potts models, the chiral Potts conditions \cite{BPAuY88}
\be \label{CPcondns}
a_p^N +k' b_p^N = k \, d_p^N \sep k' a_p^N +  b_p^N = k \, c_p^N 
\comma \ee
were immediately introduced. Here $k, k'$ are fixed constants 
(the same for all sites and edges of the lattice), satisfying
\be \label{kandk'}
k^2 + {k'}^2 = 1 \period \ee

The point of this paper is to emphasize that these conditions are 
{\em not} needed at the first step of the procedure. Even without them,
and with $a_p, b_p, c_p, d_p$ allowed to take arbitrary values for each 
of the $2L$ vertical rapidity lines of the lattice, it is still 
true that the column-to-column
transfer matrix (the ``$Q$'' matrix) of the $\tau_2 (t_q)$ model
commutes with that of the  original six-vertex model (though 
not with one another). Further, the 
row-to-row  transfer  matrices of two models $\tau_2 (t_q)$, 
$\tau_2 ({t_q}')$ (with  different horizontal rapidity variables $t_q$,
 ${t_q}'$ but the same vertical $p$-rapidities) commute. These 
$\tau_2 (t_q)$ and $\tau_j (t_q)$ matrices satisfy straightforward
generalizations of the functional relations (4.27) of 
Ref. \cite{BBP90}

We shall also show, still {\em without} the conditions 
(\ref{CPcondns}), that we can define a chiral Potts model that 
is related to the $\tau_2 (t_q)$ model by appropriate
generalizations of the transfer matrix functional relations of 
Ref. \cite{BBP90}. It is, however, in general inhomogeneous, its 
Boltzmann weights being of the usual form, but the parameters therein
being related in a rather complicated algebraic manner both to the $p$ 
variables and to the $t_q$ variable. Further, its row-to-row transfer 
matrices do not in general commute with one another, except in the 
particular combination
\bd T(\omega^i x, \omega^j y) \widehat{T} (\omega^k x,\omega^l y) \ed
defined in equations (\ref{defW}) - (\ref{elTxy}), (\ref{defhW}) -
 (\ref{That}) below.\footnote{The $\hat{T}(x,y)$ of this paper
generalizes the  $\hat{T}$ of Ref. \cite{BBP90}, but with $x$ 
and $y$ interchanged.}(Here $i, j, k, l$ are arbitrary 
integers.) Such a two-row  transfer matrix is basically that of 
the ``superintegrable'' chiral  Potts model.\cite{RJB1989}

Finally, we shall remark that this general inhomogeneous model includes
as a special case the homogeneous $\tau_2(t_q)$ model with closed
(fixed-spin) boundary conditions. The functional relations between
the  $\tau_j (t_q)$ matrices then simplify, and the eigenvalue 
spectrum is that of a direct product of $L$ single-spin matrices.
This agrees with the properties of such a superintegrable chiral 
Potts model that we observed in \cite{RJB1989}.

We have written the conditions (\ref{CPcondns}), (\ref{kandk'}) down
because they are so usually associated with the chiral Potts 
model.\footnote{They ensure that the
column-to-column transfer matrices of the $\tau_2(t_q)$ model, i.e.  
the $Q$ model, commute.}
Here we {\em never} use them. The rapidity 
$p = \{ a_p, b_p, c_p, d_p \}$ has value $p(m) = \{ a_{p(m)}, b_{p(m)}, 
c_{p(m)}, d_{p(m)} \}$ on vertical rapidity  line $m$, for 
$m = 1, 2, \ldots , 2 L$. There is no restriction on the complex 
numbers $a_{p(m)}, b_{p(m)}, c_{p(m)}, d_{p(m)}$.



\subsection*{The six-vertex  model}
We start by defining a six-vertex model in a particular field.
For this model the doubled vertical rapidity lines 
$p_1, \ldots , p_{2L}$ in Figure 
\ref{sqlattice} should be replaced by single ``type $q$'' rapidity 
lines $r_1, \ldots , r_L$.


Associate a spin $\sigma_i$ with each site $i$ of the square lattice 
$\cal L$, and allow  $\sigma_i$ to take $N$  successive integer values, 
say  $1, 2, \ldots , N$. These can be  extended  to all integer 
values with  the modular $N$ convention  $\sigma_i = \sigma_i + N$.
Not all  values are allowed: vertically adjacent spins
$\sigma_j, \sigma_k$, with $k$ above $j$ as in Figure 
\ref{sqlattice}, must satisfy the adjacency rule:
\be \label{vertadj}
\sigma_k \eq \sigma_j \; \; \; {\rm or }\; \; \; 
\sigma_j -1  \sep {\rm mod} \; \; N   \period \ee
and horizontally adjacent spins $\sigma_i, \sigma_j$, with
$j$ to the right of $i$, must satisfy
\be \label{horadj}
\sigma_j \eq \sigma_i \; \; \; {\rm or }\; \; \; 
\sigma_i + 1  \sep {\rm mod} \; \; N   \period \ee


\setlength{\unitlength}{1pt}
\begin{figure}[hbt]
\begin{picture}(420,80) (-40,0)
\multiput(20,0)(53,0){6}{\line(1,0) {30}}
\multiput(20,30)(53,0){6}{\line(1,0) {30}}
\multiput(20,0)(53,0){6}{\line(0,1) {30}}
\multiput(50,0)(53,0){6}{\line(0,1) {30}}
\put (16,-10) {$\scriptstyle{a}$}
\put (48,-10) {$\scriptstyle{a } $}
\put (16,36) {$\scriptstyle{a}$}
\put (48,36) {$\scriptstyle{a } $}
\put(33,-28) {1}
\multiput(53,0)(53,0){1}{\put (19,-10) {$\scriptstyle{a }$}
\put (43,-10) {$\scriptstyle{a+1 } $}
\put (16,36) {$\scriptstyle{a -1}$}
\put (48,36) {$\scriptstyle{a } $}
\put(33,-28) {2}}
\multiput(106,0)(53,0){1}{\put (19,-10) {$\scriptstyle{a }$}
\put (40,-10) {$\scriptstyle{a+1 } $}
\put (19,36) {$\scriptstyle{a }$}
\put (40,36) {$\scriptstyle{a+1 } $}
\put(38,-28) {3}}
\multiput(159,0)(53,0){1}{\put (19,-10) {$\scriptstyle{a }$}
\put (46,-10) {$\scriptstyle{a  } $}
\put (16,36) {$\scriptstyle{a -1 }$}
\put (41,36) {$\scriptstyle{a -1 } $}
\put(33,-28) {4}}
\multiput(212,0)(53,0){1}{\put (19,-10) {$\scriptstyle{a }$}
\put (40,-10) {$\scriptstyle{a+1 } $}
\put (21,36) {$\scriptstyle{a }$}
\put (49,36) {$\scriptstyle{a  } $}
\put(33,-28) {5}}
\multiput(265,0)(53,0){1}{\put (19,-10) {$\scriptstyle{a }$}
\put (45,-10) {$\scriptstyle{a  } $}
\put (16,36) {$\scriptstyle{a - 1 }$}
\put (49,36) {$\scriptstyle{a  } $}
\put(33,-28) {6}}
 \end{picture}
\vspace{1.5cm}
 \caption{ The six spin configurations of the six-vertex model.}
 \label{sixvertex}
\end{figure}
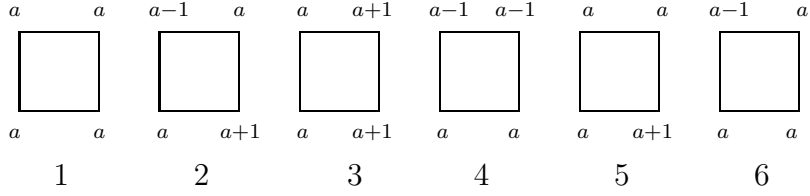

A typical face $i,j,k,l$, with the corner sites $i,j,k,l$ arranged 
anti-clockwise form the bottom-left of the square lattice is shown in 
Figure  \ref{sqlattice}. With each face $i,j,k,l$ associate a
Boltzmann weight function 
$W_{6V}(\sigma_i,\sigma_j,\sigma_k,\sigma_l)$. If $\sigma_i = a$
is fixed, then there are only six possible choices of the other three
spins, as shown in Figure \ref{sixvertex}. We define
the corresponding  Boltzmann weights to be
\ba \label{svwts}
W_{6v}(a,a,a,a) \; \; \; \;  =  \omega t -1  & , & \! \! \!
W_{6v}(a,a \plus 1,a,a \minus 1)  =  \omega t -1 \comma \nonumber \\
W_{6v}(a,a \plus 1,a \plus 1,a)  =   t -1  & , & \! \! \!  
W_{6v}(a,a,a \minus 1,a \minus 1)  =   \omega (t -1) \comma  \\
W_{6v}(a,a \plus 1,a,a) \;  =  \omega  -1  & , &\! \! \!  
W_{6v}(a,a,a,a \minus 1)  \; \;   =   \; 
(\omega  -1) \, t \comma  \nonumber \ea
for all integers $a$. Here
\be \label{defomega}
 \omega \eq \e^{2 \pi \i/N}   \ee
and  $t$ is a free parameter. For all other (non-allowed)
values of $ a, b, c, d$ we take $W_{6V}(a,b,c,d)$
to be zero. We may exhibit the $t$ dependence by writing  
$W_{6V}(a,b,c,d)$ as $W_{6V}(t|a,b,c,d)$.

The partition function is 
\be \label{partfn}
Z \eq \sum \prod_{ijkl} W_{6V}(\sigma_i,\sigma_j,\sigma_k,\sigma_l) 
\period \ee
Here the product is over all faces $(i, j, k, l )$ of the lattice, 
with cyclic (toroidal) boundary conditions. The outer sum is over 
all values of all the spins.

One can regain the usual arrow picture of the six-vertex model by
drawing arrows on the edges of the dual lattice, pointing to the left  
or up if the spins on either side are equal, to the right or down 
else. Then the six spin configurations of Figure  \ref{sixvertex}
become those of Fig. 8.2 of Ref. 
\cite{book82}, where two arrows point into each vertex and two point 
out.  Note that the weights (\ref{svwts}) are not those of the usual 
zero-field  six-vertex  model, since the weights of the third and 
fourth  configurations are 
unequal. A field with weight $\omega^{1/2}$ has been applied to
the third and fourth weights.\footnote{Pasquier and Saleur consider 
the hamiltonian associated with the six-vertex model in this special 
field.\cite{PasSal1990}} Apart from this field, 
the $\lambda, v$ of  eqn (9.2.3) of \cite{book82} are related to our 
present variables $\omega$ by $\e^{- 2\lambda} = \omega$, 
$\e^{\lambda+v} = t$.


With each horizontal (vertical) rapidity $q_i$ ($r_j$) we associate
a parameter $t_{q_i}$ $t_{r_j}$. Then the model 
is solvable if for each face 
\be t = t_q/t_{r} \comma \ee
$q = q_i$ being the horizontal rapidity and  $r = r_j$
the vertical. This is in part because the function $W_{6V}$
satisfies the star-triangle relation
\bd
\sum_g  W_{6V}(t_q|b,c,g,a )  W_{6V}(t_r|a,g,e,f )  
W_{6V}(t_r/t_q|g,c,d,e )
\eq \ed
\be \label{startri1}
\sum_g W_{6V}(t_r/t_q|a,b,g,f ) W_{6V}(t_r|b,c,d,g)  
W_{6V}(t_q|g,d,e,f)   \ee
for all $t_q, t_r$ and all values of the external spins 
$a, b, c, d, e, f$. This relation is depicted graphically in 
Figure \ref{startrifig}, provided we take $W_1, W_2, W_3$
therein to be  $W_{6V}(t_q), W_{6V}(t_r), W_{6V}(t_r/t_q)$.

\setlength{\unitlength}{1pt}
\begin{figure}[hbt]
\begin{picture}(420,80) (-40,0)
\put(40,0) {\line (1,0) {40}}
\put(20,34) {\line (1,0) {40}}
\put(80,0) {\line (3,5) {20}}
\put(20,34) {\line (3,-5) {20}}
\put(60,34) {\line (3,-5) {20}}
\put(40,68) {\line (1,0) {40}}
\put(20,34) {\line (3,5) {20}}
\put(60,34) {\line (3,5) {20}}
\put(80,68) {\line (3,-5) {20}}
\put(180,0) {\line (1,0) {40}}
\put(200,34) {\line (1,0) {40}}
\put(220,0) {\line (3,5) {20}}
\put(180,0) {\line (3,5) {20}}
\put(160,34) {\line (3,-5) {20}}
\put(180,68) {\line (1,0) {40}}
\put(160,34) {\line (3,5) {20}}
\put(220,68) {\line (3,-5) {20}}
\put(180,68) {\line (3,-5) {20}}
\put(10,31) {$a$}
\put(32,-10) {$b$}
\put(80,-10) {$c$}
\put(103,31) {$d$}
\put(80,73) {$e$}
\put(32,73) {$f$}
\put(53,26) {$g$}
\put(43,10) {$W_1$}
\put(43,47) {$W_2$}
\put(74,31) {$W_3$}
\put(150,31) {$a$}
\put(174,-10) {$b$}
\put(220,-10) {$c$}
\put(244,31) {$d$}
\put(220,73) {$e$}
\put(174,73) {$f$}
\put(201,26) {$g$}
\put(204,10) {$W_2$}
\put(204,47) {$W_1$}
\put(173,31) {$W_3$}
\put(125,31) {$=$}
 \end{picture}
\vspace{1.5cm}
 \caption{ The generalized star-triangle relation }
 \label{startrifig}
\end{figure}
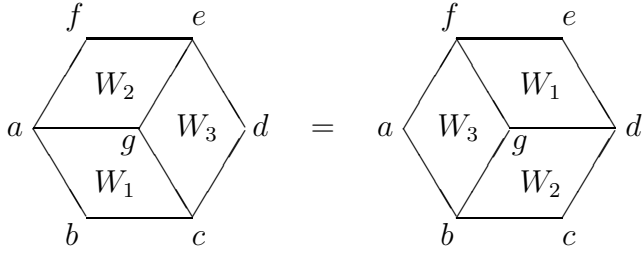

\subsubsection*{Transfer matrices}

In this $N$-state spin formulation, the row-to-row transfer matrix 
of the six vertex model is an $N^L $ by $N^L$ matrix $U_{6V}$,
with entries
\be \label{rowtm}
\left[ U_{6V} \right]_{a,b} \eq \prod_{j=1}^L W_{6V}
 (a_j, a_{j+1},b_{j+1}, b_j ) \comma \ee
where $a = \{ a_1, \ldots , a_L \}$ is the set of spins in
the lower of two successive rows of the lattice $\cal L$,
and $b = \{ b_1, \ldots , b_L \}$ is the set of spins in the 
row immediately above. It depends on $t$, so can be written
as $U_{6V}(t)$.

Similarly, the column-to-column transfer matrix is  $V_{6V}$ = 
 $V_{6V}(t)$,
where
\be \label{coltm}
\left[ V_{6V} \right]_{a,b} \eq \prod_{j=1}^M W_{6V}
 (a_j, b_j, b_{j+1},a_{j+1} ) \comma \ee
and $a = \{ a_1, \ldots , a_M \}$ is the set of spins in one column, 
$b = \{ b_1, \ldots , b_M \}$, is  the set of spins one column to 
the right, and  $M$  is the number of rows of the lattice.

Regarding  the spins $b,c,e,f$ as fixed, the star-triangle relation 
(\ref{startri1}) can be viewed as the element $(a,d)$ of an 
$N$ by $N$ matrix relation. It involves matrices with entries
$W_3 (a,c,d,e)$, $W_3 (a,b,d,f)$. Provided these are
invertible (which they usually are), the relation ensures that
\cite[\S9.6]{book82}
\be
U_{6V}(t_q) \, U_{6V}(t_r) \eq U_{6V}(t_r) \, U_{6V}(t_q)
\comma \ee
i.e. the row-to-row  transfer matrices commute, for all choices of 
$t_q, t_r$.

Similarly, regarding $a,c,d,f$ as fixed and each side of 
(\ref{startri1}) as the element $(b,e)$ of an $N$ by $N$  matrix, 
it also implies that 
\be
V_{6V}(t_q) \, V_{6V}(t_r) \eq V_{6V}(t_r) \, V_{6V}(t_q)
\comma \ee
so the column-to-column transfer matrices also commute with one 
another.



\subsection*{The $\tau_2(t_q)$ model}

The $\tau_2(t_q)$ model is also an $N$-state model on the square 
lattice $\cal L$, but now the spins only need to satisfy the
vertical adjacency rule (\ref{vertadj}). The horizontal rule
(\ref{horadj}) is {\em not} imposed. The partition function is again
given by (\ref{partfn}), but with $W_{6V}$ replaced by 
$W_{\tau}$.

The vertical adjacency rule means that the Boltzmann weight 
function $W_{\tau} (a,b,c,d)$ is zero unless $a-d = 0$ or 1 
(mod $N$) and $b-c = 0$ or 1 (mod $N$). If these constraints are 
satisfied, then 
\bd
W_{\tau} (a,b,c,d) \eq W_{\tau} (t_q |a,b,c,d)  \eq \ed
\be \label{Wtau} \sum_m \omega^{m(d-b)} (-\omega t_q)^{a-d-m} 
F_{pq}(a-d,m) F_{p' q}(b-c,m) \comma \ee
where
\bd F_{pq} (0,0 ) \; = \; 1 \sep F_{pq}(0,1)  \; = \; -\omega 
c_p t_q/b_p \comma \ed
\be \label{defF}
F_{pq} (1,0 )  \; = \; d_p/b_p \sep F_{pq}(1,1)  \; = 
\; -\omega a_p/b_p \period \ee
This function $W_{\tau}$ is the multiplicand of eqn. (3.44a) of 
Ref. \cite{BBP90}.
It is linear in the rapidity variable $t_q$ and is the Boltzmann weight 
of two triangles $\{a,d,m\}$, $\{b,c,m\}$, summed over the common spin 
$m$, with value 0 or 1, as represented in Figure \ref{bwttau2}. The 
triangles have weights $F_{pq}(a-d,m), F_{p' q}(b-c,m)$. There are 
also edge weights $\omega^{md}, \omega^{-mb}, (-\omega t_q)^{a-d}$, 
and  a site weight $(-\omega t_q)^{-m}$.

Here $p$ denotes the four complex variables $a_p, b_p, c_p, d_p$. 
Auxiliary variables that we shall use are
\be\label{defxytmu}
 x_p = a_p/d_p \sep y_p = b_p/c_p \sep t_p = x_p y_p \sep 
\mu_p  = d_p/c_p \period \ee
Similarly for $p'$. Throughout this paper we impose no restrictions
on $a_p, b_p, c_p, d_p$ (or $a_{p'}, b_{p'}, c_{p'}, d_{p'}$). 
They are {\em independent} variables.

In (\ref{Wtau}) $p$ and $p'$ are the values of $p$ for the 
particular face of the lattice under consideration. If the face is 
between spin columns $J$ and $J+1$, then
$p \eq p_{2J-1}$ and  $p' \eq p_{2J}$.

We see the reason for the doubling of the vertical 
rapidity lines in Figure \ref{sqlattice}. The odd rapidities
$p_1, p_3, \ldots , p_{2L-1}$ are those of triangles such as the one
on the left in Figure \ref{bwttau2}, with weight $F_{pq}$. The even
rapidities $p_2, p_4, \ldots , p_{2L}$ are those of triangles 
on the right in Figure \ref{bwttau2}, with weight $F_{p'q}$.

\setlength{\unitlength}{1pt}
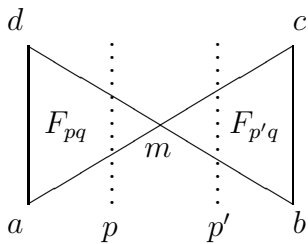
\begin{figure}[hbt]
\begin{picture}(420,100) (-70,0)
\put(80,0) {\line(5,3) {100}}
\put(80,60) {\line(5,-3) {100}}
\multiput(110,0)(0,5){13}{.}
\multiput(150,0)(0,5){13}{.}
\put(80,0) {\line(0,1) {60}}
\put(180,0) {\line(0,1) {60}}
\put (72,-11) {$a$}
\put (180,-11) {$b$}
\put (72,66) {$d$}
\put (180,66) {$c$}
\put (124,18) {$m$}
\put(108,-12) {$p$}
\put(148,-12) {$p'$}
\put(86,27) {$F_{pq}$}
\put(156,27) {$F_{p'q}$}
\end{picture}
\vspace{0.5cm}
\caption{The Boltzmann weight $W_{\tau}(a,b,c,d)$ of the $\tau_2$ model 
as that of two osculating triangles. If $a, d$ are in column $J$ and 
$b,c$ are in column $J+1$, then $p \eq p_{2J-1} $ and 
$p' \eq p_{2J} $.}
\label{bwttau2}
\end{figure}


The remarkable feature of the $\tau_2(t_q)$ model is that 
 $W_{\tau} $ and $W_{6V}$ together satisfy a second 
star-triangle relation:
\bd
\sum_g  W_{\tau}(t_q|b,c,g,a )  W_{\tau}(t_r|a,g,e,f )  
W_{6V}(t_r/t_q|g,c,d,e )
\eq \ed
\be \label{startri2}
\sum_g W_{6V}(t_r/t_q|a,b,g,f ) W_{\tau}(t_r|b,c,d,g)  
W_{\tau}(t_q|g,d,e,f)   \ee
for all $t_q, t_r$.

This can be obtained from (\ref{startri1}) simply by replacing
the first two $W_{6V}$  functions on each side by $W_{\tau}$. 
It is also represented by Figure \ref{startrifig}, but now
$W_1, W_2, W_3$ therein should be  replaced by 
$W_{\tau}(t_q), W_{\tau}(t_r), W_{6V}(t_r/t_q)$.

We can define row-to-row and column-to-column transfer
matrices $\tau_2 (t_q)$, $V_{\tau} (p, p')$ for the 
$\tau_2$ model by replacing $W_{6V}$ in (\ref{rowtm})
and (\ref{coltm}) by  $W_{\tau}$. Then 
(\ref{startri2}) implies that
\be \label{tau2comm}
\tau_2(t_q) \tau_2(t_r) \eq \tau_2(t_r) \tau_2(t_q)
\comma \ee
i.e. the row-to-row  transfer matrices commute for all $t_q$. 

Also, from (\ref{startri2}),
\be
V_{6V}(t) V_{\tau}(p,p') \eq V_{\tau}(p,p') V_{6V}(t)
\comma \ee
for all choices of $t, p, p', t_q$ in (\ref{svwts}), (\ref{Wtau}), 
(\ref{defF}). The column-to-column  transfer matrices of the 
six-vertex model therefore  commutes with that of any particular 
$\tau_2$ model. This was the  starting-point of Bazhanov
and Stroganov's derivation\cite{BazStrog90}. Note that it
does {\em not} imply that the column-to-column transfer matrices
of two different $\tau_2$ models commute. This is because when 
$\omega$ has  the particular ``root of unity'' value 
(\ref{defomega}) the eigenvalues of the six-vertex model are 
degenerate.

We shall not consider column-to-column  transfer matrices any further
herein. All the transfer matrices we shall write down in subsequent
equations will be row-to-row matrices of dimension $N^L$ by $N^L$.
The vertical $p$-rapidities are to be regarded as constants and the 
horizontal $q$-rapidity parameters $t_q$ as variables, in general 
complex.

We note in passing that a useful check on both star-triangle 
relations is provided by noting from (\ref{svwts}) that
\bd W_{6V}(1|a,b,c,d) = \delta (a,c) \period \ed
When $t_q = t_r$, it immediately follows that $g=d$ ($g=a$) on the 
LHS (RHS) of each relation, and that both are trivially satisfied.

Evaluating (\ref{Wtau}) for the four values of $a-d$ and $b-c$, we
find that in each case it is linear in the variable $t_q$, so
from (\ref{rowtm}) the matrix $\tau_2(t_q)$   
is a polynomial in $t_q$ of degree
at most $L$ (usually it is of degree $L$). {From} the commutation
relation (\ref{tau2comm}) (assuming, as seems to be the case, that 
the eigenvalues are not identically degenerate), the eigenvectors
of $\tau_2(t_q)$ must be independent of $t_q$. The  eigenvalues are 
therefore also polynomials in $t_q$ of degree $L$.



\section*{The $\tau_2, T$ relation}

The commutation relation (\ref{tau2comm}) is true
for any two $\tau_2$ models with the same vertical rapidities
$p_1, p_2, \ldots , p_{2L}$ and different horizontal 
rapidities $t_q, t_r$. Here $p_j$ is short-hand for the set
$\{ a_{p_j}, b_{p_j}, c_{p_j}, d_{p_j} \}$, so there are actually $8L$
complex numbers specifying the vertical rapidities. We emphasize 
that there are {\em no} constraints on these numbers. They can 
{\em all}  be chosen {\em independently}  and  (\ref{tau2comm})  will 
still be satisfied.

The object of this paper is to generalize the transfer matrix
functional relations of Ref. \cite{BBP90} to this arbitrary
inhomogeneous model. We start with the $\tau_2, T$ relation of 
section 4 
therein. All equation numbers herein  that contain a decimal point, 
e.g. (4.10), are references to equations of Ref. \cite{BBP90}.

Without loss of generality, we can take $k=0$ in \cite{BBP90}.
Then (4.4) becomes
\ba \label{G1}
[G_J(a)]_{m,m'}  \;  =  \! \! \! & \! \! \!\! \! \!  & \! \! \! 
  \sum_d \omega^{m' d - m a}  (-\omega t_q)^{a-d-m'} \nonumber \\
&&\; \; \; \;  F_{pq}(a \minus d,m')
 F_{p' q} (a \minus d,m)\,  g_J(d)  \ea
and (4.9) is
\be 
\sum_{m'=0} ^1 [G_J(a)]_{m,m'} \, (-r_{J+1})^{m'} \eq g'_J (a) \, 
(-r_J)^m \period \ee
Here $m, m'$ take the values $0,1$ and the sum in (\ref{G1}) is over
the allowed values $a, a-1$ of the spin $d$. 

\setlength{\unitlength}{1pt}
\begin{figure}[hbt]
\begin{picture}(420,100) (-70,0)
\put(80,30) {\line(5,3) {50}}
\put(80,30) {\line(5,-3) {50}}
\put(130,0) {\line(5,3) {50}}
\put(130,60) {\line(5,-3) {50}}
\multiput(107,0)(0,5){13}{.}
\multiput(153,0)(0,5){13}{.}
\put(130,0) {\line(0,1) {60}}
\put (122,-11) {$a$}
\put (122,66) {$d$}
\put (74,18) {$m$}
\put (174,18) {$m'$}
\put(103,-12) {$p'$}
\put(151,-12) {$p$}
\end{picture}
\vspace{0.5cm}
\caption{The sites and faces of $\cal L$ involved in the working
of section 4 of \cite{BBP90}.  If $a, d$ are in column $J$, then 
$p' \eq p_{2J-2} $ and $p \eq p_{2J-1} $.}
\label{BBP90_4}
\end{figure}
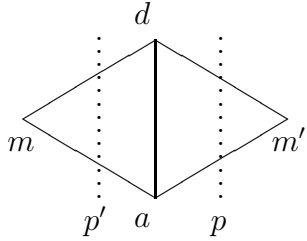

The RHS of (\ref{G1}) is the Boltzmann weight of two successive
triangles of the $\tau_2$ model, as shown in Figure \ref{BBP90_4},
with associated edge and site weights,
and an additional  site weight $g_J(d)$. The spins $a, d$ are in
column $J$ ($ J = 1, \ldots , L$) of the lattice shown in Figure 
\ref{sqlattice}, so the $p, p'$ here
are the rapidities $p_{2J-1}, p_{2J-2}$:
\be \label{pshort}
p \eq p_{2J-1} \sep p' = p_{2J-2} \period \ee

With this identification, (4.11) follows and (4.12) is
\be \label{reqns}
\frac{a_{p_{2J-2}}^N-d_{p_{2J-2}}^N r_J^N} {c_{p_{2J-2}}^N 
t_q^N -b_{p_{2J-2}}^N r_J^N} 
\times \frac{d_{p_{2J-1}}^N t_q^N -a_{p_{2J-1}}^N r_{J+1}^N} 
{b_{p_{2J-1}}^N  -c_{p_{2J-1}}^N r_{J+1}^N} 
\eq 1 \ee
for $J = 1, \ldots , L$, taking $p_0 = p_{2L}$.
This is the condition that the function $g_J(a)$ be 
periodic of period $N$, i.e. $g_J(a+N) = g_J(a)$. We need this because
the spins in the $\tau_2$ model only take $N$ values and are always 
to be interpreted as integers modulo $N$.

Given $r_1^N$, we can solve the bilinear relation (\ref{reqns}) 
successively
for $r_2^N, \ldots ,$  $r_L^N$, $r_{L+1}^N$. Since $r_{L+1} = r_1$, 
this gives
a quadratic relation for  $r_1^N$, and hence for all the $r_J^N$.
In \cite{BBP90}, we then used the fact that we were taking the vertical
rapidities to be those of the usual chiral Potts model to obtain the 
two explicit solutions (4.13) for  the $r_J^N$.

Here we can no longer do this: all we can say is that the sequence
$r_1^N, \ldots , r_L^N$ is one of the two solutions of
(\ref{reqns}).\footnote{They do have some interesting properties, as 
in equation (\ref{alphaalt}).}

We can still carry on with the rest of the working. In place of (4.13) 
we set
\be \label{defxy}
r_J \eq \omega^{1-\beta_{J-1}} \, \x_{J-1}  \sep \y_{J-1} \eq 
t_q/\x_{J-1}  \comma \ee
where $\beta_{J-1}$ is an integer that does not enter the relations 
(\ref{reqns}). For given vertical rapidities $p_1, \ldots , p_L$,
we take $t_q, \x_1, \ldots , \x_L, \y_1, \ldots , \y_L$ 
to be fixed, satisfying (\ref{reqns}) and  (\ref{defxy}). Then we allow
$\beta_1, \ldots , \beta_L$ to take any set of integer values.

Equations (4.15) - (4.18) still follow, provided we replace 
$\omega^{-\beta_{J}} a_q/d_q$ by  $\omega^{-\beta_{J}} \x_{J}$,
$\omega^{-\beta_{J}} c_q/b_q$ by  $\omega^{-\beta_{J}}/ \y_{J}$,
and similarly with $J$ replaced by $J-1$. Then in place of (4.19) 
we obtain
\bd g_J (a) \eq y_p y_{p'} \,  \overline{W}_{J-1}(a \minus 
\beta_{J-1}|\omega \x_{J-1},\y_{J-1})
\, W_{J}(a \minus \beta_{J}|\omega \x_{J},\y_{J}) \comma \ed

\bd g'_J (a) \eq \frac{(y_{p}-\omega \x_J) (t_{p'} - t_q) }
{x_{p'} - \x_{J-1}} \,  \overline{W}_{J-1}(a \minus 
\beta_{J-1}|\x_{J-1},\y_{J-1})
\, W_{J}(a \minus \beta_{J}|\x_{J},\y_{J}) \comma \ed

\bd g''_J (a) \eq \frac{(1-y_{p'}/\y_{J \minus 1}) 
(t_{p} - \omega t_q) }
{1-x_p /\y_J} \,  \overline{W}_{J-1}(a \minus 
\beta_{J-1}|\omega \x_{J-1},\omega \y_{J-1}) \ed \bd 
\times \, W_{J}(a \minus \beta_{J}|\omega \x_{J}, \omega 
\y_{J}) \period \ed
Here $p = p_{2J-1}$,  $p' = p_{2J-2}$ and $x_p, y_p, t_p$, 
$x_{p'}, y_{p'}, t_{p'}$ are defined by (\ref{defxytmu}).
 The functions 
$W$, $\overline{W}$ are given by
\be \label{defW}
 W_{J}(a |x_{J},y_{J}) \eq \prod_{i=1}^a \frac{d_{p_{2J-1}}  - 
\omega^i  a_{p_{2J-1}}/y_J }{b_{p_{2J-1}}  - \omega^i 
c_{p_{2J-1}} x_J } \comma \ee

\be  \label{defWb}
\overline{W}_{J}(a |x_{J},y_{J}) \eq \prod_{i=1}^a \frac{\omega 
a_{p_{2J}}  - \omega^i  d_{p_{2J}} x_J }{c_{p_{2J}}  - \omega^i 
b_{p_{2J}}/ y_J } \period \ee

Note the distinction between $x$ and $y$ with a $p$ or $p'$ suffix, and 
$x$ and $y$ with an integer $J$ or $J-1$ suffix. The former are 
defined by  (\ref{defxytmu}) and are vertical rapidity variables. The 
latter are defined by (\ref{defxy}). In fact,
$\x_J, \y_J$ are generalizations of the $x_q, y_q$ of \cite{BBP90}, 
so can be thought of as ``$q$ variables'', but they also 
depend via (\ref{reqns}) on  all the vertical rapidities.

Thes functions are generalizations of the chiral Potts model 
edge-weight
functions $W$, $\overline{W}$ \cite[eqns. 2 and 3]{BPAuY88}, 
\cite[eqns. (2.4) and (2.5)]{RJB93}. Their definitions can be 
extended
to negative integers $a$ in the usual way:
\bd \prod_{i=1}^a s_i \eq \prod_{i=a+1}^0 1/s_i  \ed
for any $s_i$.
However, they do {\em not} satisfy the usual periodicity conditions
$W(a+N) = W(a)$, $\overline{W}(a+N) = \overline{W}(a)$. Instead
they satisfy the weaker condition
\be \label{wcond} 
\frac{W_J(a+N |  x_J,   y_J)
\overline{W}_{J-1}(b+N |  x'_{J-1},  y'_{J-1})}
{W_J(a |  x_J,   y_J)
\overline{W}_{J-1}(b |  x'_{J-1},  y'_{J-1})} \eq 1  \ee
where $x'_J = \omega^i x_J$,  $y'_J = \omega^i y_J$ 
and the  integers $a,b,i,j$ are arbitrary. In fact this is just the 
condition (\ref{reqns}). It ensures that the functions $g_J(a)$, 
$g'_J(a)$, $g''_J(a)$ are all periodic of period $N$.

\subsubsection*{The ``chiral Potts'' transfer matrix}
We now generalize the usual definition of the chiral  Potts
transfer matrix in (2.15a) and define a matrix $T(x,y)$
with entries
\be \label{elTxy}
T(x,y)_{a,\beta} \eq \prod_{J=1}^L
W_J(a_J-\beta_J| x_J, y_J) \overline{W}_{J - 1}(a_J - \beta_{J
\minus 1}| x_{J \minus 1}, y_{J \minus 1}) \period \ee
Here $x = \{ x_1, \ldots , x_L \}$,  $y = \{ y_1, \ldots , y_L \}$,  
 $a = \{ a_1, \ldots , a_L \}$ and  $\beta = \{ \beta_1, \ldots ,
 \beta_L \}$.  

Because of (\ref{wcond}), incrementing any of the spins 
$a_1, \ldots , a_L $ by $N$ leaves $T(x,y)_{a,\beta} $ unchanged. Thus
the rows of the matrix $T(x,y)$ have the same modulo-$N$ spin 
invariance property as the rows and columns of the 
$\tau_2$-model tranfer matrix $\tau_2(t_q)$. Restricting each of these
spins to $N$ values,  $\tau_2(t_q)$ is a square $N^L$ by $N^L$ matrix;
$T(x,y)$ has $N^L$ rows.

The columns of $T(x,y)$, labelled by $\beta_1, \ldots, \beta_L$,
 are slightly more subtle. Incrementing any $\beta_J$ by $N$ does 
change  the RHS of (\ref{elTxy}), but only by multiplying it by a 
factor independent of $a_1, \ldots , a_L$. Further, this factor 
depends on 
$x_1, \ldots , y_L$ only via their $N$th powers. It follows 
that  $T(x,y)$ has at most $N^L$ linearly independent columns, and 
numerical calculations strongly suggest that in general there are
indeed $N^L$ linearly independent columns. Thus although we may take 
$T(x,y)$ to have more than  $N^L$ columns, there is a unique
$N^L$ by $N^L$ matrix  $S_{ij}(x,y) $ such that
\be \label{Sij}
S_{ij}(x,y) \;  T(x,y) \eq T(\omega^i x,\omega^j y) \ee
for all integers $i, j$. We formally write $S_{ij}(x,y) $
as
\bd
 T(\omega^i x,\omega^j y) \, T(x,y)^{-1} \period \ed

With these definitions, eqn. (4.20) of \cite{BBP90} becomes
\be \label{4.20a}
\tau_2(t_q) \, T(\omega x,y) \eq c(x,t_q) \, T(x,y) + d(y,t_q) \, 
T(\omega x, \omega y) \comma \ee
where
\bd
c(x,t_q) \eq \prod_{J=1}^L \frac{(y_{p_{2J-1}}-\omega \, x_J) 
(t_{p_{2J}}-t_q)}
{y_{p_{2J-1}} \, y_{p_{2J}} \, (x_{p_{2J}}-x_{J})} \comma \ed
\be \label{defd}
d(y,t_q) \eq \prod_{J=1}^L \frac{(y_{p_{2J}}- y_{J}) 
(t_{p_{2J-1}}- \omega t_q)}
{y_{p_{2J-1}} \, y_{p_{2J}} \, (x_{p_{2J-1}}-y_{J})} \period \ee

As in (2.42), define an $N^L$ by $N^L$  matrix  $X$ with entries
\be
X_{\sigma, \sigma'} \eq \prod_{J=1}^L \delta(\sigma_J, \sigma'_J+1) 
\period \ee
This is the operator that shifts all spins in a row by one. 
It commutes with $\tau_2(t_q)$:
\be
X \, \tau_2(t_q) \eq \tau_2(t_q) \,  X \period \ee

Replacing $x_J, y_J$ in (\ref{defW}), (\ref{defWb}) by 
$\omega^{-1} x_J, \omega y_J$ is equivalent to replacing the index
$i$ by $i-1$. It follows that
\be
T(\omega^{-1} x, \omega y) \eq \rho(x,y) \, X \, T(x,y) \comma \ee
 where
\be
\rho(x,y) \eq \prod_{J=1}^L \frac{\mu _{p_{2J-1}} \mu_{p_{2J}} 
(1-x_{p_{2J-1}}/y_J) (\omega \, x_{p_{2J}}-x_J)}
{ (y_{p_{2J-1}}-x_J) (1- y_{p_{2J}}/y_J)} \period \ee

One other function that we need is given by the obvious generalization 
of (4.23):
\be \label{defz}
z(t_q) \eq \prod_{J=1}^{L} \omega \, \mu_{p_{2J-1}} \mu_{p_{2J}} 
(t_{p_{2J-1}}-t_q) (t_{p_{2J}}-t_q) /
(y_{p_{2J-1}} y_{p_{2J}})^2  \period \ee
It is a polynomial in $t_q$, of degree $2L$.

Then one can verify that
\be c(x,\omega t_q) d(y,t_q) \rho(\omega x,y) \eq z(\omega t_q)  \ee
from which it follows that we write (\ref{4.20a}) as
\be \label{4.20b}
\tau_2(t_q) \, T(\omega x,y) \eq \frac{z(t_q)}
{d(\omega^{-1}y, \omega^{-1} t_q)} \, X \, T(\omega x,\omega^{-1} y) 
+ d(y,t_q) \, 
T(\omega x, \omega y) \period \ee



\subsection*{The $\tau_2, \widehat{T}$ relation}

We obtained (\ref{4.20b}) by considering an $N^L$-dimensional vector
$g$ which is a direct product of $L$ vectors of dimension $N$,
forming $\tau_2(t_q) g$ and finding the conditions under which
this is the sum of two such direct product vectors. There are
$N^L$ linearly independent vectors $g$. The matrix with these columns 
is $T(x,y)$.

We can also form instead the row-vector $g^T \, \tau_2(t_q)$. 
Corresponding working goes through and we are led to define
two more Boltzmann weight edge-functions
\be \label{defhW}
 \widehat{W}_{J}(a |x_{J},y_{J}) \eq \prod_{i=1}^a \frac{\omega 
\, d_{p_{2J}} x_J  - 
\omega^i  \, a_{p_{2J}} }{b_{p_{2J}}/y_J  - \omega^{i-1} 
c_{p_{2J}}  } \comma \ee
\be  \label{defhWb}
\widehat{\overline{W}_{J}}(a |x_{J},y_{J}) \eq 
\prod_{i=1}^a \frac{\omega 
\, a_{p_{2J-1}}/y_{J}  - \omega^{i-1}  \, d_{p_{2J-1}}  }{\omega 
c_{p_{2J-1}} x_J   - \omega^i \, b_{p_{2J-1}} }  \ee
and a transfer matrix $\widehat{T}(x,y)$ with entries
\be \label{That}
\widehat{T}(x,y)_{\beta a} \eq \prod_{J=1}^L 
\widehat{\overline{W}_{J}}(\beta_J -a_J |x_{J},y_{J}) 
 \widehat{W}_{J-1}(\beta_{J-1} - a_{J} |x_{J-1},y_{J-1}) \period \ee

Setting 
\be \widehat{d}(y,t_q) \eq \prod_{J=1}^L 
\frac{\omega (y_J - y_{p_{2J}} )
(t_{p_{2J-1}} - t_q) }{y_{p_{2J-1}} y_{p_{2J}} 
(y_J - \omega \, x_{p_{2J-1}})} \comma \ee
we obtain the analogue of (4.21):
\be \label{4.21a}
\widehat{T}(x,\omega y) \, \tau_2(t_q) \eq \frac{z(\omega t_q) }
{\widehat{d}(\omega y, \omega t_q) } \, \widehat{T} (x,\omega^2 y) X +
\widehat{d}(y,t_q) \, \widehat{T}(x,y) \period \ee



\section*{The $\tau_j$ relations}

Here we extend the $\tau_2$ matrices to the set 
$\tau_1, \ldots , \tau_{N+1}$, where $\tau_j = \tau_j(t_q)$
is a polynomial in $t_q$ of degree $(j-1) L$. We shall not give  
explicit definitions in terms of lattice models, as is done 
in section 3 of Ref. \cite{BBP90}, but will use only the relation 
(\ref{4.20b}).

We start by defining
\ba 
\Delta_r(x,y) = && \! \! \! \! \! \! d(y,t_q) 
d(\omega y, \omega t_q) \cdots  
d(\omega^{r-1} y, \omega^{r-1} t_q)  \sep  r  \geq 0 \nonumber \\
  = &&  \! \! \! \! \! \!  \{
d(\omega^{-1} y,\omega^{-1} t_q) d(\omega^{-2} y, \omega^{-2} t_q) 
\cdots   d(\omega^{r} y, \omega^{r} t_q) \}^{-1} \sep   r \leq 0 
\comma \ea
\be \label{defD}
D_r(x,y) \eq \Delta(x,y) \,  T(\omega x, \omega^r y)
T(\omega x,  y)^{-1} \period \ee
In particular, $D_0(x,y) = \I$ is the identity matrix. {From}
(\ref{Sij}), we expect the RHS to exist and be unique.

The relation (\ref{4.20b}) can then be written
\be \label{4.20d}
\tau_2(t_q) \eq z(t_q) X D_{-1}(x,y) + D_1(x,y) \period \ee
The spin-shift operator $X$ commutes with $\tau_2(t_q)$ and with
all the $D_r(x,y)$. It is consistent with (\ref{4.20d}) that all the
matrices $D_r(\omega^i x,\omega^j y)$ commute with $\tau_2(t_q)$
and with one another, for all $r, i, j$. This is what we observe 
numerically: we
shall assume that this is so.

We now define  $\tau_j(t_q)$ to be
\be \label{deftauj}
\tau_j(t_q) \eq \sum_{k=0}^{j-1} z(t_q) z(\omega t_q) \cdots 
z(\omega^{k-1} t_q) \, X^k \, D_k(x,\omega^{k-1} y) \, 
D_{j-k-1} (x,\omega^k y) 
 \ee
for $j \geq 0$. Then
\bd \tau_0(t_q) = 0 \sep  \tau_1(t_q) = \I \ed 
and $\tau_2(t_q)$ is as given in (\ref{4.20d}). {From} this definition 
it follows that
\newcounter{newctr}
\setcounter{newctr}{\value{equation}}
\addtocounter{newctr}{1}
\setcounter{equation}{0}
\renewcommand{\theequation}{\thenewctr \alph{equation}}
\be \label{4.27a}
\tau_2(\omega^{j-1} t_q ) \tau_j (t_q) \eq z(\omega^{j-1} t_q) X \, 
\tau_{j-1}(t_q) + \tau_{j+1}(t_q) \comma \ee
\be \label{4.27b}
\tau_j(\omega t_q ) \tau_2 (t_q) \eq z(\omega t_q) X \, 
\tau_{j-1}(\omega^2 t_q) + \tau_{j+1}(t_q) \comma \ee
\be \label{4.27c}
\tau_{N+1} (t_q) \eq z(t_q) X \tau_{N-1} (\omega t_q) + \alpha_q 
+\overline{\alpha}_q \comma \ee
\setcounter{equation}{\value{newctr}}
\renewcommand{\theequation}{\arabic{equation}}
for $j=1, \ldots, N$, where 
\be \label{defalpha}
\alpha_q \eq \prod_{i=0}^{N-1}  d(\omega^i y,\omega^i t_q)  
\comma \ee
\be \label{defalphabar}
\overline{\alpha}_q \eq \prod_{i=0}^{N-1} z(\omega^i t_q)/
d(\omega^i y,\omega^i t_q)  \period \ee
The two sets of relations (\ref{4.27a}), (\ref{4.27b}) are 
equivalent. Note that $\alpha_q, \overline{\alpha}_q$ are
unchanged by the mapping
\bd x, y, t_q \rightarrow x, \omega y, \omega t_q \period \ed

These equations are the generalizations of (4.27) - (4.29). 
In deriving them we have kept  
$x = \{ x_1, \ldots , x_L \}$ fixed and incorporated all 
multiplications by powers of $\omega$ into 
$y= \{ y_1, \ldots , y_L \}$ and therefore $t_q$, so 
if we write the rhs of (\ref{deftauj}) more explicitly as
$\tau_j(x,y)$, then by $\tau_j(\omega^k t_q)$ in the above equations 
we mean $\tau_j(x,\omega^k y)$. However, (\ref{4.27a}) or 
(\ref{4.27b}) can be used to successively form 
$\tau_3(t_q)$, $\tau_4(t_q)$, etc. Since $z(t_q)$ and 
$\tau_2(t_q)$  are polynomials in $t_q$ of degree $2L$ $L$, 
respectively, from this construction it 
follows that each $\tau_j(t_q)$ is also a polynomial in $t_q$, of 
degree $(j-1) L$. So $\tau_j(t_q)$ is indeed a single-valued function
of $t_q$, unchanged by replacing $x,y$ by $\omega x, \omega^{-1} y$,
and by the choice of the  solution of (\ref{reqns}).

{From} (\ref{tau2comm}) it follows that all the matrices $\tau_j(t_q)$
commute, for all values of $t_q$. There is therefore a similarity
transformation, independent of  $t_q$, that simultaneously
diagonalizes all the $\tau_j(t_q)$. Then (\ref{4.27a}) - 
(\ref{4.27c}) become scalar functional relations for each eigenvalue,
which is also a polynomial in $t_q$. These relations define the 
eigenvalue. There are many solutions, corresponding to the different 
eigenvalues.

Another way of looking at this is to note that if we replace $y$ 
in (\ref{4.20b}) by $\omega y, \omega^2 y, \ldots, \omega^{N-1} y$,
we obtain a total of $N$ homogeneous linear equations for 
$N$ unknowns $T(\omega x,y), \ldots , T(\omega x,\omega^{N-1} y)$.
The determinant of these relations must vanish, and that
is the relation for the function $\tau_2(t_q)$ obtained by eliminating
$\tau_3(t_q), \ldots , \tau_{N+1}(t_q)$ from (\ref{4.27a}) and 
(\ref{4.27c}), or equivalently from (\ref{4.27b}) and 
(\ref{4.27c}).

We have derived the hierarchy of relations (\ref{4.27a}) - 
(\ref{4.27c}) from (\ref{4.20b}). We could equally well have 
derived them from (\ref{4.21a}).



\subsection*{Calculation of $\alpha_q + \overline{\alpha}_q$}

Since each matrix function $\tau_j(t_q)$ is a polynomial in $t_q$,
from (\ref{4.27c}), the same must be true of 
$\alpha_q + \overline{\alpha}_q$, and it must be of degree at most 
$N L$.
This is by no means obvious: it appears from (\ref{defalpha}) and 
(\ref{defalphabar}) that $\alpha_q$ and $\overline{\alpha}_q$ are 
each quite
complicated functions of the solution $x_1, \ldots , x_L$ of 
(\ref{reqns}). The object of this section is to unravel this 
little mystery.

{From} (\ref{defxytmu} and (\ref{defxy}), using the shorthand notation
(\ref{pshort}), we can write the the condition (\ref{reqns}) as
\be \label{23a}
\mu_p^N \mu_{p'}^N \frac{(x_{p'}^N - x_{J-1}^N) (t_q^N- x_p^N x_J^N)}
{(t_q^N - y_{p'}^N x_{J-1}^N )( y_p^N - x_J^N) } \eq 1 \period \ee
We can use this relation to eliminate the factors containing
$y_J$ in  (\ref{defd}) and (\ref{defalpha}). Using also  (\ref{defz})
and  (\ref{defalphabar}),  it follows that  
\ba \label{alalbar}
\alpha_q & = &  \prod_{J=1}^L \frac{\mu_p^N \mu_{p'}^N 
\, (x_{p'}^N-x_{J-1}^N ) \, (t_p^N-t_q^N) }{y_p^N y_{p'}^N \, 
(y_p^N-x_J^N)} \nonumber \\
 \overline{\alpha}_q & = & \; \; \; \; \; \prod_{J=1}^L \frac{
\, (y_{p}^N-x_{J}^N ) \, (t_{p'}^N-t_q^N) }{y_p^N y_{p'}^N \, 
(x_{p'}^N-x_{J-1}^N)} \comma \ea
where in the multiplicands we again write $p, p'$
for  $p_{2J-1},  p_{2J-2}$, respectively.

Since $J = 1, \ldots , L$ and $x_0 = x_L$, (\ref{23a}) is a set of 
 $L$ equations for $x_1^N, \ldots ,x_L^N$. We noted above that it 
has two solutions. Let the other solution be
 ${x'_1}^N, \ldots, {x'_L}^N$. Then from (\ref{23a})
\be \label{firsteqn}
 (x_{p'}^N - x_{J-1}^N) (t_q^N- x_p^N x_J^N)
(t_q^N - y_{p'}^N {x'}_{J-1}^N )( y_p^N - {x'_J}^N) \eq \ee
\bd
(x_{p'}^N - {x'}_{J-1}^N) (t_q^N- x_p^N {x'_J}^N)
(t_q^N - y_{p'}^N x_{J-1}^N )( y_p^N - x_J^N) \period \ed

This equation can be re-written in the ``Wronskian'' form:
\bd
(x_{J-1}^N - {x'}_{J-1}^N) (t_q^N - x_p^N x_J^N) (y_p^N - {x'}_J^N)
(t_{p'}^N-t_q^N) \eq \ed
\be
(x_{J}^N - {x'}_{J}^N) (t_q^N - y_{p'}^N x_{J-1}^N) (x_{p'}^N - 
{x'}_{J-1}^N)
(t_{p}^N-t_q^N) \period \ee

We can use (\ref{23a}) to eliminate the factors 
$(t_q^N - x_p^N x_J^N)$, $(t_q^N - y_{p'}^N x_{J-1}^N) $, leaving
\ba  \frac{(y_p^N-x_J^N)  (t_{p'}^N-t_q^N)}{y_p^N 
y_{p'}^N (x_{p'}^N-x_{J-1}^N)} \brk  \eq \nonumber \\ 
\brk \close  \frac{\mu_p^N \mu_{p'}^N (x_J^N-{x'}_J^N) 
(x_{p'}^N- {x'}_{J-1}^N) 
(t_p^N-t_q^N) }{ y_p^N y_{p'}^N (x_{J-1}^N-{x'}_{J-1}^N) 
(y_p^N-{x'}_J^N) } \period \ea
Taking the product over $J = 1, \ldots , L$, the factors
$(x_J^N-{x'}_J^N)$, $(x_{J-1}^N-{x'}_{J-1}^N)$ cancel, giving
\be \label{alphaalt}
\overline{\alpha}_q \eq \left[ \alpha_q \right]' \comma \ee
where $\left[ \alpha_q \right]'$ is defined by
(\ref{alalbar}), but with each $x_J$ replaced by $x'_J$.
Interchanging each $x_J$, $x'_J$, it follows at once that
$\alpha_q = \left[ \overline{\alpha}_q \right]'$, so
$\alpha_q + \overline{\alpha}_q$ is {\em unchanged} by
replacing the solution $x_1, \ldots, x_L$ by the alternative
solution $x'_1, \ldots, x'_L$. It is therefore a single-valued 
function of $t_q$.

Now we look at (\ref{23a}), considered as a recursion relation
giving $x_{J-1}^N$ in terms of $x_J^N$. Set
\be x_J^N  = f_J/g_J \ee
for $J=1, \ldots , L$. Then we can choose the normalization
so that
\ba \label{recurr}
-y_p^N y_{p'}^N f_{J-1} &  = & (t_q^N - \mu_p^N \mu_{p'}^N 
x_p^N x_{p'}^N ) f_J +
(\mu_p^N \mu_{p'}^N x_{p'}^N - y_{p'}^N) t_q^N g_J \nonumber \\
-y_p^N y_{p'}^N g_{J-1} & = & 
(y_{p'}^N - \mu_p^N \mu_{p'}^N x_p^N  ) f_J +
(\mu_p^N \mu_{p'}^N t_{q}^N - y_{p}^N y_{p'}^N) g_J \comma \ea
where again $p = p_{2J-1}$, $p' = p_{2J-2}$. This is a 
linear relation
for $(f_{J-1},g_{J-1})$ in terms of $(f_J,g_J)$.

With these definitions, we find that the multiplicand in the
second equation (\ref{alalbar}) is simply $g_{J-1}/g_J$, so 
\be \label{simpalpha}
\overline{\alpha}_q \eq \prod_{J=1}^L 
g_{J-1} /g_J    \eq g_0/g_L \period \ee

\renewcommand{\arraystretch}{1.3}

Define two-by-two matrices
\bd 
A_{2J} \eq \left( \begin{array}{cc} t_q^N & \mu_{p_{2J}}^N 
x_{p_{2J}}^N  \\
y_{p_{2J}}^N & \mu_{p_{2J}}^N \end{array} \right)/y_{p_{2J}}^N 
\comma \ed
\be \label{defAB}
B_{2J-1} \eq \left( \begin{array}{cc} -1 &  y_{p_{2J-1}}^N  \\
 \mu_{p_{2J-1}}^N x_{p_{2J-1}}^N & -\mu_{p_{2J-1}}^N t_q^N 
\end{array} \right)
/y_{p_{2J-1}}^N  \ee
\renewcommand{\arraystretch}{1.0}
and set
\bd \xi_J \eq \left( \begin{array}{c} f_J  \\
g_J \end{array} \right) \period \ed
Then  (\ref{recurr}) can be written as
\be \xi_{J-1} \eq A_{2J-2} B_{2J-1} \xi_J \period \ee
Since $x_0^N = x_L^N$, it follows that $\xi_0 = \lambda \xi_L$,
where 
\be \label{eigvaleqn}
\xi_0 = \lambda \, \xi_L = A_{2L} B_1 A_2 B_3 \cdots A_{2L-2} 
B_{2L-1}  \, \xi_L \ee
Thus $\lambda$ is the eigenvalue of 
$U =  A_{2L} B_1  \cdots A_{2L-2} B_{2L-1}$, $\xi_L$ is the 
corresponding eigenvector and, from (\ref{simpalpha}),
\be \overline{\alpha}_q \eq \lambda \period \ee

Since $U$ is a two-by-two matrix, it has two eigenvalues
$\lambda$ and $\lambda'$, corresponding to the two solutions
$x$ and $x'$ of the recurrence relations. However, we have just shown 
that interchanging the solutions replaces  $ \overline{\alpha}_q$
by $\alpha_q$, so
\be \label{resalphabar}
\alpha_q \eq \lambda' \period \ee

Since $\lambda + \lambda'$ is the trace of the matrix $U$, 
it follows that 
\be \label{alphasum}
\alpha_q + \overline{\alpha}_q   \eq 
{\rm Trace} \; \left( A_{2L}  B_1 A_2 B_3 \cdots A_{2L-2}
 B_{2L-1} \right) \period \ee

We are regarding the vertical rapidity parameters 
$x_{p_1}, y_{p_1},\mu_{p_1}, \ldots ,$  $x_{p_L}$,$y_{p_L}$, 
$\mu_{p_L}$ 
as constants and  $t_q$ as a complex variable, so this 
is an explicit expression for $\alpha_q + \overline{\alpha}_q $
that makes it clear that it is indeed a polynomial in $t_q^N$.
Since $A_{2J-2} B_{2J-1} $ is linear in $t_q^N$,  this polynomial 
is of degree not greater than $L$ (in general it is of degree $L$).

{From} (\ref{defz}), (\ref{alalbar}) and (\ref{defAB}) it is readily 
seen that  
\bd \lambda \lambda' = {\rm det} \, U = \prod_{i=1}^N z(\omega^i t_q)
 = \alpha_q \overline{\alpha}_q \comma \ed
so we could have obained (\ref{resalphabar}) without going
through the working from equation (\ref{firsteqn}) to 
(\ref{alphaalt}). We have included that working, partly for 
completeness, but also  because it is an elegant example of how 
in solvable models the algebra conspires to produce needed results.

The $\tau_j(t_q)$ relations (\ref{4.27a}) - (\ref{4.27c}), together 
with (\ref{defAB}) and (\ref{alphasum}), provide a closed set of
equations that determine the eigenvalues of the $\tau_j(t_q)$
matrices, all quantities being polynomials in the complex variable 
$t_q$. To use them, there is no need to solve the eigenvalue
equation (\ref{eigvaleqn}), which is equivalent to the recurrence 
relation (\ref{reqns}). We could presumably have obtained these 
relations
directly by a ``fusion'' method, generalizing the definition 
(3.26) - (3.44) of $\tau_j(t_q)$, but this is quite technical.
We prefer the present approach, based on the equation (\ref{4.20b}).

We have assumed that the matrices $D_r(\omega^i x, \omega^j y)$
commute with one another and with $\tau_2(t_q)$. This 
assumption agrees with numerical calculations we have performed
for $N = L = 3$, but it can probably be removed. We can 
certainly apply a 
similarity transformation (independent of $t_q$) that diagonalizes
$\tau_2(t_q)$ (for all $t_q$). Applying this only to the left of 
(\ref{4.20b}), it becomes a set of many equations for each eigenvalue
of $\tau_2(t_q)$. If we focus on just one eigenvalue and one
such equation,
then, as we remarked above, we can obtain $N$ relations from it
by replacing $y, t_q$ by $\omega^i, \omega^i t_q$, for $i=0,
\ldots , N-1$. These are homogeneous linear relations for 
the corresponding $N$ elements of $T(\omega x, \omega^i y)$, 
so their determinant must vanish. The resulting 
determinantal relation
must be equivalent to (\ref{4.27a}) - (\ref{4.27c}).






\subsection*{The $T, \widehat{T}$ relations.}

{From} the definitions (\ref{deftauj}) we can also establish that
\bd
\alpha_q \tau_j(t_q) + z(t_q) \cdots z(\omega^{j-1} t_q) 
X^j \tau_{N-j} (\omega^j t_q) \eq
D_j(x, \omega^{-1} y) \,  \tau_N (t_q) \ed
for $j = 0, \ldots , N$. {From} (\ref{defD}), there must therefore be 
a matrix $Y(x,y)$, independent of $j$,  such that
\be \label{hier}
\Delta_j(x,\omega^{-1} y) T(\omega x, \omega^{j-1}  y) Y(x,y) \eq
\alpha_q \tau_j(t_q) + z(t_q) \cdots z(\omega^{j-1} t_q) 
X^j \tau_{N-j} (\omega^j t_q) \ee
for $j = 0, \ldots , N$.

These equations have the same structure as the fusion hierarchy
of relations (3.46), except we still have to identify the matrix
$Y$. We have not fully done this, but we can note from 
(\ref{defW}), (\ref{defWb}), (\ref{defhW}), (\ref{defhWb}) 
that
\be \label{Wparrl} 
\overline{W}_J(a|\omega x_J,y_J) \,  \widehat{W}_J(-a |x_J,y_J) 
= 1 \comma \ee
\ba \label{Wser}
\sum_c W_J(a-c|\omega x_J, y_J) \! \! \! \! && \! \! \! \!
\widehat{\overline{W}}_J
(c-b |x_J, y_J )  =  \nonumber \\
&& \! \! \! \! \! \! \! \!  \! \! \! \! \! \! \! \! 
\frac{N (x_p \minus \omega^{-1} y_J) (y_p \minus 
\omega x_J) (t_p^N \minus t_q^N) }{(t_p \minus t_q)  (x_p^N 
\minus y_J^N)  (y_p^N \minus x_J^N) } \, \delta_{a,b} \comma \ea
where $p = p_{2J-1}$, and $\delta_{a,b} = 1$ if $a = b$ to  modulo
 $N$, else
$\delta_{a,b} = 0$. The sum is over any $N$ consecutive integer
values of $c$: although the $W$ functions individually are
not periodic functions, the products in the above two equations are
indeed periodic functions of $a, b, c$, of period $N$.

Consider the matrix product $T(\omega x, y) \widehat{T}(x,y)$,
by which we mean the usual sum over the intermediate  
indices, in this case both being the $\beta$ indices
in (\ref{elTxy}), (\ref{That}). Incrementing any $\beta_J$ by
$N$ multiplies the columns of  $T(\omega x, y)$ by certain factors,
but divides the rows of $\widehat{T}(x,y)$ by the same factors,
so leaves their product unchanged. Thus we can naturally take 
the intermediate 
sum to be over the values $0, \ldots , N \minus 1$ 
(or any set of $N$ successive values) of each of $\beta_1, \ldots,
\beta_{L}$.

\setlength{\unitlength}{1pt}
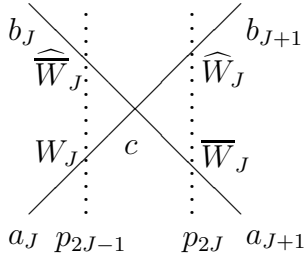
\begin{figure}[hbt]
\begin{picture}(420,100) (-70,0)
\put(80,0) {\line(1,1) {80}}
\put(80,80) {\line(1,-1) {80}}
\multiput(100,0)(0,5){16}{.}
\multiput(140,0)(0,5){16}{.}
\put (72,-11) {$a_J$}
\put (162,-11) {$a_{J+1}$}
\put (72,66) {$b_J$}
\put (162,66) {$b_{J+1}$}
\put (116,23) {$c$}
\put(90,-12) {$p_{2J-1}$}
\put(138,-12) {$p_{2J}$}
\put (82,20) {$W_J$}
\put (145,18) {$\overline{W}_J$}
\put (82,49) {$\widehat{\overline{W}}_J$}
\put (145,50) {$\widehat{W}_J$}
\end{picture}
\vspace{0.5cm}
\caption{The Boltzmann weight of a face of the 
transfer matrix product $T(\omega x, y) \widehat{T}(x,y)$.}
\label{ThatT}
\end{figure}

The matrix product is the Boltzmann weight of two successive rows
of the lattice. A typical face of this double row is shown in
Figure \ref{ThatT}. Suppose the two external spins 
$a_{j+1}, b_{J+1}$ are equal. Then from (\ref{Wparrl} ), the 
Boltzmann weight factors $\overline{W}_J \widehat{W}_J$ factors
cancel. The sum over the centre spin $c$ then gives the RHS of
(\ref{Wser}), which vanishes unless $a_J = b_J$, mod $N$.

If  $a_J = b_J$, then the same argument applied to the face to 
the left tells us that $a_{J-1} = b_{J-1}$, and so on. It follows
that
\be \label{invn}
T(\omega x, y) \widehat{T}(x,y) \eq \left[ \, \tau_1 (t_q) + R \, 
\right] /g(x,y) \comma \ee
where $R$ is a matrix with non-zero elements $R_{ab}$ only when 
$a_1 \neq b_1 $, $\! \! \ldots \, $, $a_L \neq b_L$, and 
\be
g(x,y) \eq \prod_{J=1}^L \frac{(t_p \minus t_q)  (x_p^N 
\minus y_J^N)  (y_p^N \minus x_J^N) }{N 
(x_p \minus \omega^{-1} y_J) (y_p \minus 
\omega x_J) (t_p^N \minus t_q^N) } \comma \ee
where each $p$ in the multiplicand is $p_{2J-1}$.

This is an ``inversion identity'': it has the same structure as the
$j=1$ case of (\ref{hier}), with the matrix $T(\omega x, y)$ on the 
left and the first term on the right 
being proportional to $\tau_1(t_q)$, i.e. to the identity matrix.
We conjecture (in agreement with numerical calculations)
that the right-hand sides of the two relations are 
in fact the same, to within a scalar factor, in which case
$R = z(t_q) X \tau_{N-1}(\omega t_q)/ \alpha_q$
and 
\be  \Delta_1(x,\omega^{-1} y)  \, Y(x,y) \eq \alpha_q \, 
g(x,y) \, \widehat{T}(x,y)  \period \ee

Then (\ref{hier}) becomes
\ba \label{3.46}
g(x,y) \,  \Delta_{j-1}(x,y)  \brk  T(\omega x, \omega^{j-1} y)
 \, \widehat{T} (x,y)
\eq   \nonumber \\
\brk  \tau_j(t_q) + z(t_q) \cdots z(\omega^{j-1} t_q) 
X^j \tau_{N-j} (\omega^j t_q) /\alpha_q  \period  \ea
This is the generalization of (3.46).
 
\subsubsection*{Consistency}

An interesting consistency check on (\ref{3.46}) is provided
by post-multiplying (\ref{4.20b}) by $\widehat{T}(x,\omega y)$,
and pre-multiplying (\ref{4.21a}) by $T(\omega x,y)$. The left-hand 
sides are then the same, equating the right-hand sides gives
\bd
\frac{z(t_q)}
{d(\omega^{-1}y, \omega^{-1} t_q)} \, X \, T(\omega x,\omega^{-1} y) 
\widehat{T}(x,\omega y) + d(y,t_q) \, 
T(\omega x, \omega y) \widehat{T}(x,\omega y) \eq \ed
\be \label{TThat_cons}
\frac{z(\omega t_q) }
{\widehat{d}(\omega y, \omega t_q) } \, T(\omega x,y) \widehat{T} 
(x,\omega^2 y) X +
\widehat{d}(y,t_q) \,T(\omega x,y)  \widehat{T}(x,y) \period \ee

We can use (\ref{3.46}) to express each of the $T \widehat{T}$
products as a sum of  $\tau_j$ terms. In fact we get only terms
proportional to $\tau_1(t_q)$, $\tau_1(\omega t_q)$,
 $\tau_{N-1}(\omega t_q)$ and  $\tau_{N-1}(\omega^2 t_q)$.
There are two terms proportional to each of these
four factors. Since $\tau_1(t_q) = \I$, we can interchange 
 $\tau_1(t_q)$ with $\tau_1(\omega t_q)$ on the RHS. 
Using only the relations
\bd  \Delta_{N-2} (x,\omega y)  \eq \alpha_q  / \{ d(y,t_q) \,
 d(\omega^{-1} y, \omega^{-1} t_q) \} \,  
\comma \ed
\bd \widehat{d} (y,t_q) \eq d(y,t_q) g(x,y)/ g(x,\omega y)  \ed
and the commutation of $X$ with all the $\tau_j$ matrices, we 
then find that the two terms for each $\tau$ factor cancel, 
thereby verifying (\ref{TThat_cons}).



\section*{The $\tau_2(t_q)$ model with open boundaries.	}

We return to considering the hierarchy of relations (\ref{4.27a}) - 
(\ref{4.27c}) for the $\tau_j(t_q)$ functions.

These relations simplify greatly when we impose fixed-spin boundary 
conditions on the left and right sides of the lattice. We can do this 
by taking 
\be \label{open}
a_{p_1} \eq d_{p_1} \eq 0 \period \ee
Then $F_{{p_1} q} (1,m) = 0$, so the weight function of Figure
\ref{bwttau2} vanishes for the faces between column 1 and column 2
unless $a = d$. This is equivalent to requiring that all the spins
in  column 1 of Figure \ref{sqlattice} be equal. The model is 
unchanged
by incrementing every spin by one, so we can in particular
require that every spin on column 1 be zero. It is evident from 
Figure \ref{sqlattice} that this is the same as requiring that all 
spins on the left and right boundaries be zero.

(\ref{open}) implies that
\be 
\mu_{p_1} \eq x_{p_1} \eq z(t_q) \eq  \overline{\alpha}_q \eq 0 \ee
so the relations (\ref{4.27a}) - (\ref{4.27c}) simplify
to
\be \label{tauprod}
\tau_2(t_q) \tau_2(\omega t_q) \cdots  \tau_2(\omega^{N-1} t_q) \eq 
\alpha \period \ee
{From} (\ref{alphasum}), noting that the second row of the matrix 
$B_1$ is now zero,
\be \label{alphasimp}
\alpha = [B_1 A_2 B_3 A_4 \cdots B_{2L-1} A_{2L} ]_{11} \period \ee
The RHS is a polynomial in $t_q^N$ of degree $L$ and we noted above 
that $\tau_2(t_q)$ and its eigenvalues are polynomials in $t_q$
of degree $L$. Further, when either $t_q$ is large or small, to 
leading order $\tau_2(t_q)$ is diagonal, with entries
\bd
\prod_{J = 1}^L (1-\omega^{a_{J+1}-a_J +1} t_q/y_{2J-1} y_{2J} ) \ed
in row and column $a = \{ a_1, \ldots , a_L \}$.
Let the  zeros of (\ref{alphasimp})
be  $s_1^N, s_2^N$,  $\ldots$,  $s_L^N$. Then it follows that all 
eigenvalues  of $\tau_2(t_q)$ are of the form
\be \label{eigform}
\Lambda(t_q) \eq (\omega^L/Y)  \,  \prod_{j=1}^L (s_j - 
\omega^{\gamma_j} t_q) \comma \ee
where 
$s_1 s_2 \cdots s_L = \omega^{-L} Y$, and 
\be Y \eq \prod_{J=1}^L y_{p_{2J-1}} y_{p_{2J}} \period \ee

The $\gamma_1, \ldots , \gamma_L$ are integers
with values in the range $0, \ldots ,N \minus 1$. They satisfy 
the condition $\gamma_1+ \cdots + \gamma_L = 0$, and it seems from 
low-temperature expansions that the full set of $N^{L-1}$ eigenvalues
is obtained by allowing $\gamma_1, \ldots , \gamma_L$ to take all 
such values (distinct to modulo $N$).

It appears that the other values of $\gamma_1, \ldots , \gamma_L$
that do  not satisfy the sum rule also correspond to eigenvalues of 
$\tau_2(t_q)$, provided 
we generalize the model to allow the skewed boundary conditions
\bd a_{L+1} = a_1 + r \ed
in every row of the lattice ( so all spins in column 1 
are still the same, as are all spins in column $L+1$, but now those
in the two boundary columns no longer need be equal). Then
\bd \gamma_1+ \cdots + \gamma_L = r \period \ed
We can take $r$ to be an integer in the range $0, \ldots , N-1$.

The eigenvalues therefore have the same simple structure as do direct 
products 
of $L$ matrices, each of size $N$ by $N$. For $N=2$ this is
the structure of the eigenvalues of the Ising 
model.\cite{Onsager1944}

For the Ising model this property follows from Kaufman's
solution in terms of spinor operators \cite{Kaufman1949},
i.e. a Clifford algebra.\cite[p.189]{Lounesto1997}
Whether there is some generalization of such spinor operators
to handle the $\tau_2(t_q)$ model with open boundaries remains a
fascinating speculation.\cite{RJBZN}

The results of this section were obtained in Ref. \cite{RJB1989}.
There we considered the superintegable chiral Potts model and 
rotated it though $90^{\circ}$ to obtain a model that is in
fact the present $\tau_{N}(t_q)$ model. Then we inverted its
row-to-row transfer matrix,
thereby obtaining the present $\tau_2(t_q)$ model. We did in fact 
note in  section 7 of \cite{RJB1989} that we could allow the modulus
$k$ to be different for different rows: this corresponds to
our here allowing $a_p, b_p,  c_p, d_p$ to all vary arbitrarily
from column to column.



\section*{Summary	}

We have shown that the column-inhomogeneous $\tau_2(t_q)$ model is
solvable for {\em all} values of the $8L$ parameters 
$a_{p_1}, b_{p_1}, c_{p_1}, d_{p_1}, \ldots , d_{p_{2L}}$, where 
$a_{p_J}, b_{p_J}, c_{p_J}, d_{p_J}$
are associated with the $J$th vertical dotted line in Figure
\ref{sqlattice}. They do {\em not} need to satisfy the
``chiral Potts'' conditions (\ref{CPcondns}). The model then
has the unusual property that its row-to-row transfer matrices
(with different values of $t_q$ but the same 
$a_{p_1}, \ldots , d_{p_{2L}}$) commute, while the column-to-column
transfer matrices do not.

Our results (\ref{4.20b}), (\ref{4.21a}), (\ref{4.27a}) - 
(\ref{4.27c}),
(\ref{3.46}) generalize the relations (4.20), (4.21), (4.27a) - 
(4.27c),
(3.46) of Ref. \cite{BBP90}. The last generalization (\ref{3.46})
is essentially a conjecture, depending as it does on the 
identification
of (\ref{invn}) with the $j=1$ case of (\ref{hier}). However, 
it has been tested numerically for $N = L = 3$ with arbitrarily
chosen values of the parameters and found to be true to the 30 
digits of precision used. 

One significant difference from the homogeneous model
is that the associated chiral Potts model weights (\ref{defW}),
(\ref{defWb}), (\ref{defhW}), (\ref{defhWb}) depend on $t_q$ via the 
solution $r_1, \ldots, r_{L}$ of (\ref{reqns}). If we change $t_q$
then we change $r_1, \ldots, r_{L}$ in a non-trivial way, so it makes
little sense to combine $T(x,y)$ for one value of $t_q$ with 
$\hat{T}(x,y)$ for another value. It appears that our
generalized chiral Potts transfer matrices $T$ and $\hat{T}$, with 
different values of $t_q$, do not satisfy any general commutation 
relations like (2.31) - (2.33) of Ref. \cite{BBP90}.

In short, we can generalize the $\tau_2(t_q)$ model to arbitrary
$a_{p_1}, \ldots , d_{p_{2L}}$, but the only chiral Potts model
we can correspondingly generalize is the ``superintegrable''
model with the alternate row-to-row transfer matrices 
$T(x,y)$, $\hat{T}(x,y)$ defined above. In each double row
$t_q, r_1^N, \ldots ,r_L^N$ must be the same for  $T(x,y)$
and  $\hat{T}(x,y)$.

The  functional relations (\ref{4.27a}) - (\ref{4.27c}) define the 
eigenvalues of the row-to-row transfer matrix $\tau_2(t_q)$. For 
fixed-spin conditions on the left and right boundaries these can be 
solved  explicitly, giving the simple ``direct product'' result
(\ref{eigform}). 


\end{document}